The Statistical Mechanical Model of Sediment Transport Capacity and Scour-and-Silt volume in Wide and Shallow Rivers

Liu Kejing[1,2]
1. State Key Laboratory of Hydroscience and Engineering, Tsinghua University, Beijing, China
2. Jimei University, Fujian, China

**Abstract**

**Purpose** This study aims to develop a universal, parameter-free model for sediment transport and riverbed evolution using a rigorous statistical physics framework. It seeks to overcome the limitations of traditional deterministic and empirical approaches by establishing formulas with general applicability.

**Methods** The river channel is conceptualized as an isothermal-isobaric ensemble containing numerous non-identical sediment particles. The macroscopic state of the system, defined by the scour-and-silt volume, is derived from the statistical mechanics of particle distributions. The Gibbs free energy and partition function for the ensemble are formulated, considering the two primary states of particles (suspended load and bed load) and the transitions between them. This theoretical framework yields a universal formula for the number of particles in transport and the consequent volumetric change.

**Results** The model was applied to six reaches of the Lower Yellow River from 2000-2001. Calculations revealed a seasonal pattern in the number of transported particles, higher in winter and lower in summer. The results showed an alternation between scour (January-July) and siltation (July-January), with a net scour volume over the 24-month period. The magnitude of scour-and-silt volume decreased from upstream to downstream, findings that are consistent with independent observational records following the operation of the Xiaolangdi Reservoir.

**Conclusion** The model successfully simulates riverbed evolution without empirical parameters, demonstrating that statistical physics provides a robust framework for predicting complex fluvial processes. Its general formulation suggests potential applicability to other similar multi-particle systems.

## 1 Introduction

Traditional physical models for river dynamics and fluvial process are established based on classical mechanics and deterministic mathematical methods (Pähtz T, et al. 2020; Wu 2024). Both sides of these equations represent processes at the same moment, i.e., "calculating the present dependent variables from the present independent variables." This grants them a complete form and solvability within the Newtonian mechanics framework, but in reality, they cannot fully describe natural processes. This shortcoming cannot be rectified within the scope of classical mechanics, moreover, such methods are limited in their ability to handle the systems with numerous particles. For a long time, the field of sediment transport has addressed this issue using various empirical formulas and semi-empirical, semi-physical models (Andualem, et al. 2023; Pähtz T, et al. 2020). Relationships developed based on one study area require parameter adjustment and modification when applied to other regions, hence there are few general models with complete physical significance. Some models, derived from long-term engineering experience, have yielded formulations capable of describing how variables evolve over time (Wu 2008). The introduction of kinetic theory has provided a theoretical foundation for describing the dynamics of individual and collective sediment particles, while also incorporating sediment transport mechanics into the domain of statistical mechanics (Wu 2024; Zhong, et al. 2015). Existing kinetic-based research has focused on characterizing the movement of sediment particles. This study attempts to employ the statistical mechanics framework to establish general models for sediment transport capacity and scour-and-silt volume in wide and shallow river channel, providing universal formulas that require no parameter calibration.

**2 Methods**

From the perspective of statistical mechanics, the state of a macroscopic physical quantity at any given moment must be one of all its statistically possible states (Pathria RK & Beale PD, 2021). Considering sediment particles as particles, their distribution under hydraulic driving forces at each moment constitutes the riverbed state at that moment, and cumulative quantities reflect the cumulative distribution of the particles.

For a wide, shallow, straight river reach within the same climate zone, where water temperature, air pressure, and water pressure do not undergo drastic changes over a period, neglecting biological actions and chemical reactions, sediment particles are treated as homogeneous, non-identical solid particles with consistent density and mechanical properties, but not necessarily identical volume and mass. It is assumed that collisions occur between sediment particles; there are two types of sediment particles: suspended load and bed load; exchange exists between these two types; and particles move within the volume bounded by the water surface and the riverbed. Let the scour-and-silt volume of an alluvial river reach at time $t$ be the macroscopic quantity. For a reach of length $L$, and under the condition of no sediment input from tributaries, aeolian sand, or other sources, this volume essentially represents the net increase or decrease of sediment particles located on the riverbed during transport within the reach over that period. The river reach is treated as an isothermal-isobaric ensemble.

In statistical mechanics, for an isothermal-isobaric process, the Gibbs free energy (Pathria RK & Beale PD, 2021) is:

$$G = \mu N = E - TS + PV \qquad (1)$$

where $\mu$ is the chemical potential (Pathria RK & Beale PD, 2021), interpreted in this study as the influence of particle number $N$ on energy, $N$ is the number of particles participating in transport,

$$\mu = kT \ln\left[\frac{N}{V}\left(\frac{h^2}{2\pi \overline{m} kT}\right)^{\frac{3}{2}}\right] \qquad (2);$$

$V$ is the system volume (m³). $P$ is the system pressure (hPa). In the absence of rapid air pressure changes (e.g., passage of strong convective weather systems) or artificial manipulation of water pressure, the water surface is considered the equilibrium surface between the atmosphere and the water body, and the value is taken as the average air pressure over the period. Boltzmann constant $k = 1.380649 \times 10^{-23}$ J/K, Planck constant $h = 6.62607015 \times 10^{-34}$ J·s, $T$ is the system temperature in thermodynamic scale (K), taken here as the average water temperature over the study period (for wide, shallow rivers where width is much greater than depth, and depth is only a few meters, the temperature difference from below the water surface to the bed is usually small). $\bar{m}$ is the average particle mass (kg), which can be calculated by $\bar{m} = \frac{4}{3}\pi \bar{r}^3 \rho_s$ ($\bar{r}$ is the mean particle radius or specified grain size, m; $\rho_s$ is particle density, which can be taken as the dry density of sediment) (文献). $S$ is the entropy (Pathria RK & Beale PD, 2021),

$$S = Nk \ln\left(\frac{V}{N}\right) + \frac{3}{2}Nk\left[\frac{5}{3} + \ln\left(\frac{2\pi \bar{m}kT}{h^2}\right)\right] \quad (3);$$

$E$ is the total system energy. The energy distribution is (Pathria RK & Beale PD, 2021):

$$p(E) = \sum_{\{s|E_s=E\}} p_s = \frac{1}{Z}\sum_{\{s|E_s=E\}} e^{-\beta E_s} = \frac{1}{Z}\Omega(E)e^{-\beta E} \quad (4)$$

where $\Omega(E)$ is the number of possible particle states corresponding to energy, $Z$ is called the partition function, representing the total number of system states, i.e., "how many possible system states exist statistically." Of course, any current state is one of the possible states. From the Boltzmann relation (Pathria RK & Beale PD, 2021)

$$\frac{\Delta S}{\Delta(\ln\Omega)} = \frac{1}{\beta T} = k \quad (5)$$

Planck (Pathria RK & Beale PD, 2021) derived
$$S = k \ln \Omega \quad (6)$$
namely
$$\Omega(E) = e^{\frac{S}{\beta T}} \quad (7)$$

Substituting Eq. (7) into Eq. (4) yields

$$p(E) = \frac{1}{Z}e^{\beta TS}e^{-\beta E} = \frac{1}{Z}e^{\beta(TS-E)} \quad (8)$$

All sediment particles participating in transport have two possible states: suspended in water and moving with the water, i.e., suspended load; or moving on the riverbed driven by water flow and concentration gradient forces, i.e., bed load. With only these two states and possible transitions between them, for two cross-sections of the river reach, any single sediment particle has four possible transition states: 1. water → water, 2. water → bed, 3. bed → water, 4. bed → bed. When $N$ particles independently exist in these four states, there are $2^{2N}$ possible states between the two cross-sections. Each state corresponds to a possible energy state. For the total system energy $E$, bed load and suspended load represent different energy levels. A change in energy indicates scour (more particles transitioning from bed load to suspended load) or siltation (more particles transitioning from suspended load to bed load) within the volume between the two cross-sections.

Among the four possible states for any sediment particle, states 1 and 4 involve no energy change, assuming that there are only two sediment particles with unequal masses between two cross-sections, among the 16 states, there are 4 states with no energy change. It is easy to know that for two cross-sections containing $N$ grains of sand, there are $3 \cdot 2^{2(N-1)}$ states with energy changes. Combining this with the definition of the partition function, we obtain the total number of system states in a river channel containing only water and sediment particles, where the changes in the states of particles are reflected as changes in the volume of the river channel., i.e., a concise expression for $Z$

$$Z = 3 \cdot 2^{2(N-1)} \qquad (9)$$

Then

$$p(E) = \frac{1}{3 \cdot 2^{2(N-1)}} e^{\beta(TS-E)} \qquad (10)$$

For a system where sediment particles have two states (in water and on the bed), when particles are located on the bed, they alter the system volume. If we set $N = N_S + N_B$, where $N_S$ is the number of suspended particles, $N_B$ is the number of bed load particles, $p(E) = p(N_B) + p(N_S)$, $p(N_B)$ represents the proportion of bed load among all sediment, $p(N_S)$ represents the proportion of suspended load among all sediment, no forms other than bedload and suspended load, thus $p(E) = 1$. From Eq. (10):

$$E = TS - \frac{1}{kT} \ln 3 \cdot 2^{2(N-1)} \qquad (11)$$

Bed load, driven by water flow under gravity, randomly accumulates and moves, forming moving accumulations on the riverbed. Assuming the reach has an original base volume, $V_0$, which is not scoured away by flow in the short term, then the system volume at any moment is this part minus the average volume of the accumulation body, $V_B$, i.e.,

$$V = V_0 - V_B \qquad (12)$$

The average volume of the accumulation body, $V_B$, is the macroscopic quantity of the microscopic structure $b$ of the particle accumulation. For a bed surface with $N$ sediment particles, there are $2^N$ states, thus

$$V_B = \frac{1}{2^N} \sum b \cdot 2^N \qquad (13)$$

Then

$$V = V_0 - \frac{1}{2^N} \sum b \cdot 2^N \qquad (14)$$

From another perspective, the riverbed consists of $m$ consecutive, extremely thin cross-sections of equal thickness, each with thickness $\Delta L = \frac{L}{m}$, where $L$ is the reach length. Each cross-section $i$ has an original base area $A_{0i}$ that is not subject to short-term scour-induced change, and an accumulation cross-sectional area $A_{Bi}$ on the cross-section. Then the system volume is:

$$\overline{V} = (A_{01} - A_{B1}) \cdot \Delta L + \cdots + (A_{0m} - A_{B_m}) \cdot \Delta L \qquad (15)$$

Transforming this equation gives:

$$\begin{aligned}
\overline{V} &= (A_{01} - A_{B1}) \cdot \Delta L + \cdots + (A_{0m} - A_{B_m}) \cdot \Delta L \\
&= \frac{(A_{01} - A_{B1}) + \cdots + (A_{0m} - A_{Bm})}{m} \cdot L \\
&= \frac{\sum_{i=1}^{m} A_{0i}}{m} \cdot L - \frac{\sum_{i=1}^{m} A_{Bi}}{m} \cdot L \\
&= \overline{A_0} \cdot L - \frac{\sum_{i=1}^{m} A_{Bi}}{m} \cdot L
\end{aligned} \qquad (16)$$

where the first term on the right side, $\overline{A_0} \cdot L \approx V_0$, and the second term on the right side is also the average volume of the moving accumulation, thus $\overline{V} \approx V$. And

$$\frac{\sum_{i=1}^{m} (A_{0i} - A_{Bi})}{m} = \overline{A}$$ is the average cross-sectional area of the reach, so $V = \overline{A} \cdot L$ can serve as an approximation for the system volume.

Substituting (2), (3), and (11) into (1) yields an equation concerning $N$:

$$\frac{5}{2} TNk - T \left\{ Nk \ln\left(\frac{V}{N}\right) + \frac{3}{2} Nk \left[\frac{5}{3} + \ln\left(\frac{2\pi \overline{m} kT}{h^2}\right)\right]\right\} + \frac{1}{kT} \ln\left[3 \cdot 2^{2(N-1)}\right] = PV \qquad (17)$$

Let $X = \frac{2\pi \overline{m} kT}{h^2}$, $Y = \frac{1}{kT} \ln 4 - kT \ln\left(V \cdot X^{\frac{3}{2}}\right)$, rearranging gives:

$$kTN \ln N + YN = PV - \frac{1}{kT} \ln \frac{3}{4} \qquad (18)$$

This is a transcendental equation concerning $N$, solved using Lambert W function, yielding:

$$N = \exp\left\{ W\left[ \left( \frac{PV}{kT} - \frac{1}{k^2T^2} \ln \frac{3}{4} \right) e^{\frac{Y}{kT}} \right] - \frac{Y}{kT} \right\} \qquad (19)$$

Here, $W(\cdot)$ is the Lambert function, which is the inverse function of $f(W) = We^W$, and $\cdot$ is any complex number. $W(\cdot)$ cannot be expressed using elementary functions and is multi-valued, but it can be approximated using an asymptotic expansion. Obviously $\left( \frac{PV}{kT} - \frac{1}{k^2T^2} \ln \frac{3}{4} \right) e^{\frac{Y}{kT}} \gg 0$, which is a very large number. Taking its 0 branch $W_0$, and from $W_0(x) = \ln x - \ln \ln x + o(1)$:

$$W_0\left[ \left( \frac{PV}{kT} - \frac{1}{k^2T^2} \ln \frac{3}{4} \right) e^{\frac{Y}{kT}} \right] = \ln\left[ \left( \frac{PV}{kT} - \frac{1}{k^2T^2} \ln \frac{3}{4} \right) e^{\frac{Y}{kT}} \right] - \ln \ln\left[ \left( \frac{PV}{kT} - \frac{1}{k^2T^2} \ln \frac{3}{4} \right) e^{\frac{Y}{kT}} \right] + o(1)$$
(20)

Substituting this into (21), and letting $Y_2 = PV - \frac{1}{kT} \ln \frac{3}{4}$, gives:

$$N = \frac{Y_2}{Y + kT \ln Y_2 - kT \ln kT} \qquad (21)$$

This is the theoretical calculation form for the total number of particles potentially participating in transport within the river reach.

Substituting Eq. (21) into Eq. (13), if $b$ is known, the average volume of the bed load accumulation at the calculation time can be obtained. The volumetric scour-and-silt amount over a specific period can be calculated by:

$$\Delta V = V_B(t_2) - V_B(t_1) \qquad (22)$$

If $\Delta V > 0$, it indicates siltation occurred in the reach during that period; if $\Delta V < 0$, it indicates scour occurred.

For $b$, we consider it as a packing problem of non-identical irregular particles. There are gaps between the packed particles. Small particles may occupy the gaps between large particles, and small particles are certainly located in the gaps of large particles. Therefore, the space occupied by a unit volume of the packing is the sum of the total volume of the particles and the void volume. Eq. (15) shows that the average volume of the packing is the average of the packing volume over all possible states. Here, various forms of particles are considered according to their equivalent-volume spheres. For $N$ particles, each with a radius of $r_i$, $i = 1, \cdots, N$ and the volume fraction of

particles in a unit volume of the accumulation is $\phi_{RCP}$, then from Eq. (13):

$$V_B = \frac{2\pi \sum_{i=1}^{N} r_i^3}{3\phi_{RCP}} \qquad (23)$$

Replacing $r_i$ with the more readily available equivalent spherical diameter of the sediment particle $D_i = 2r_i$, this equation can be written as:

$$V_B = \frac{\pi \sum_{i=1}^{N} D_i^3}{12\phi_{RCP}} \qquad (23b)$$

$\phi_{RCP}$ can be determined by simulating the free accumulation of particles with various size distributions. According to research by Desmond KW & Weeks ER (2014), $\phi_{RCP} = \phi_{RCP}^* + c_1\delta + c_2 s\delta^2$, where $\phi_{RCP}^* = 0.634$, $c_1 = 0.0658$, $c_2 = 0.0857$, the polydispersity $\delta = \sqrt{\langle\Delta r^2\rangle}/\langle r\rangle$, the skewness $s = \langle\Delta r^3\rangle/\langle\Delta r^2\rangle^{3/2}$, $\langle r\rangle$ is the mean of radius, $\Delta r = r - \langle r\rangle$. Substituting Eq. (25b) into (24) gives:

$$\Delta V = \frac{\pi}{12}\left[\frac{\sum_{i=1}^{N(t_2)} D_i^3}{\phi_{RCP}(t_2)} - \frac{\sum_{i=1}^{N(t_1)} D_i^3}{\phi_{RCP}(t_1)}\right] \qquad (24)$$

This is the calculation model for the volumetric scour-and-silt amount.

**3 Simulation and Calculation**

Taking the Lower Yellow River as an example, eight hydrological stations (Xiaolangdi, Huayuankou, Jiahetan, Gaocun, Sunkou, Aishan, Luokou, Lijin) are distributed along the relatively straight downstream channel. Based on the channel measurement data (Han 2025) in the following two tables, the monthly scour-and-silt volumes for the reaches Huayuankou-Jiahetan (HJ), Jiahetan-Gaocun (JG), Gaocun-Sunkou (GS), Sunkou-Aishan (SA), Aishan-Luokou (AL), and Luokou-Lijin (LL) for the years 2000-2001 are calculated.

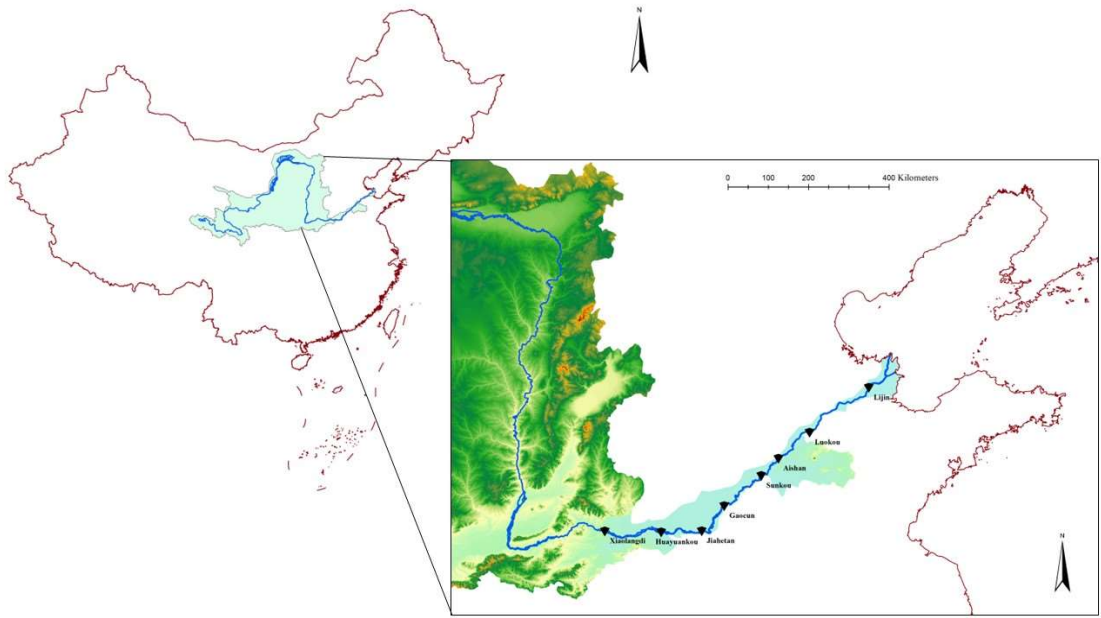

Figure 1 Study Area and Station Locations

Table 1 Sediment Gradation in River Reaches

| Boundary particle size (mm) | Reach | | | | | |
| --- | --- | --- | --- | --- | --- | --- |
| | HJ | JG | GS | SA | AL | LL |
| | Propotion | | | | | |
| 0.000002 | 0.007 | 0.0046 | 0.0062 | 0.0059 | 0.0118 | 0.0125 |
| 0.000004 | 0.004 | 0.0043 | 0 | 0.0033 | 0.0075 | 0.0135 |
| 0.000008 | 0.009 | 0.0049 | 0.0089 | 0.007 | 0.0155 | 0.017 |
| 0.000016 | 0.009 | 0.0077 | 0.0115 | 0.0117 | 0.022 | 0.022 |
| 0.000031 | 0.016 | 0.0135 | 0.017 | 0.0146 | 0.0373 | 0.035 |
| 0.000062 | 0.049 | 0.05 | 0.077 | 0.1065 | 0.126 | 0.125 |
| 0.000125 | 0.1668 | 0.289 | 0.3685 | 0.4475 | 0.419 | 0.415 |
| 0.00025 | 0.4145 | 0.478 | 0.433 | 0.371 | 0.3419 | 0.34 |
| 0.0005 | 0.2927 | 0.14 | 0.075 | 0.0325 | 0.019 | 0.02 |
| 0.001 | 0.032 | 0.008 | 0.0029 | 0 | 0 | 0 |

Table 2 Geometric Characteristics of Reaches and Dry Bulk Density of Sediment

| River Reach | Main Channel Cross-Sectional Area (km²) | Reach Length (km) | Dry Bulk Density of Sediment (t/m³) |
| --- | --- | --- | --- |
| HJ | 469.85 | 100.8 | 1.51 |
| JG | 299.48 | 72.6 | 1.51 |
| GS | 221.15 | 118.2 | 1.47 |
| SA | 59.91 | 63.9 | 1.47 |
| AL | 63.74 | 101.8 | 1.47 |
| LL | 109.75 | 167.8 | 1.47 |

The water temperature was not obtained, and was calculated using the formula

$$t_w = 4.717\left[\frac{(1+\theta^2)^{0.0781}}{(1+0.325\omega^2)^{0.0325}}\right]e^{0.041t}$$ proposed by Li, et al (2006), here, $\theta$ is the relative humidity, $\omega$ is the surface wind speed (m/s), and $t$ is the air temperature (°C). The meteorological data comes from ECMWF ERA5 (Hersbach, et al. 2023).

The average particle volume was obtained from the boundary particle sizes (diameters) and the proportions of each size fraction in Table 1, and then the average particle mass was derived using the dry bulk density (density) from Table 2.

The calculated monthly number of sediment particles participating in transport for the six reaches in 2000-2001 is shown in Figure 2. The monthly scour-and-silt volumes along the course are shown in Figure 3. The total scour-and-silt volumes along the course for the 24 months from 2000-2001 are shown in Figure 4.

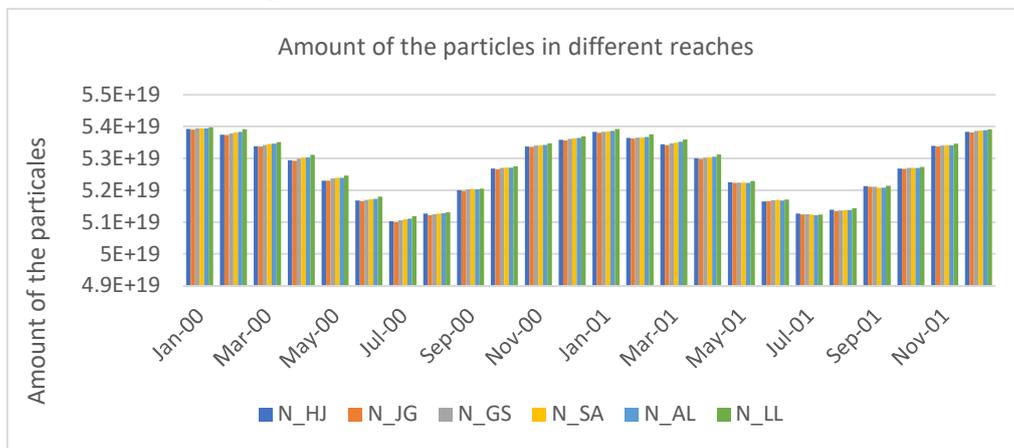

Figure 2. Monthly Number of Sediment Particles Participating in Transport in the Huayuankou-Lijin Reach of the Lower Yellow River, 2000-2001

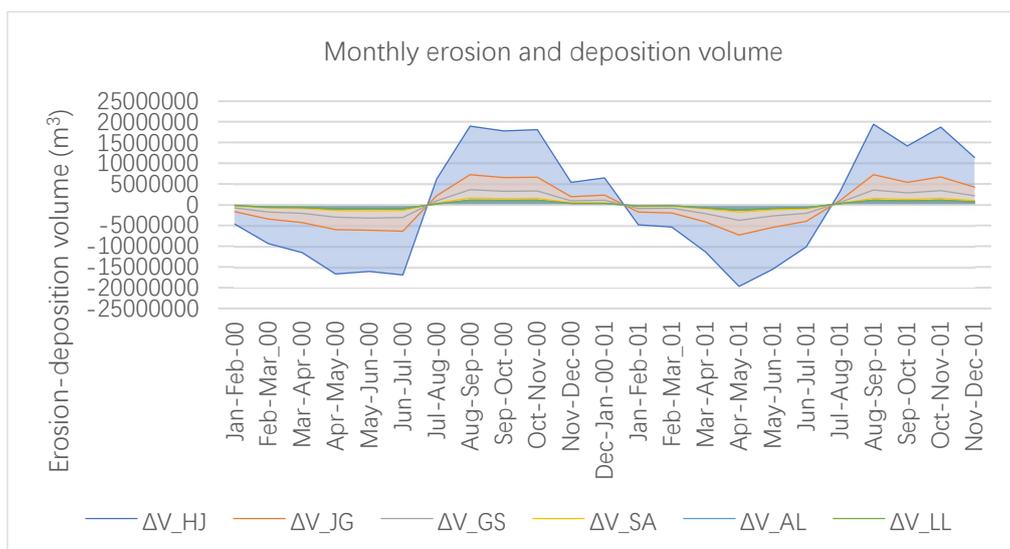

Figure 3 Monthly Scour-and-Silt Volume by River Reach

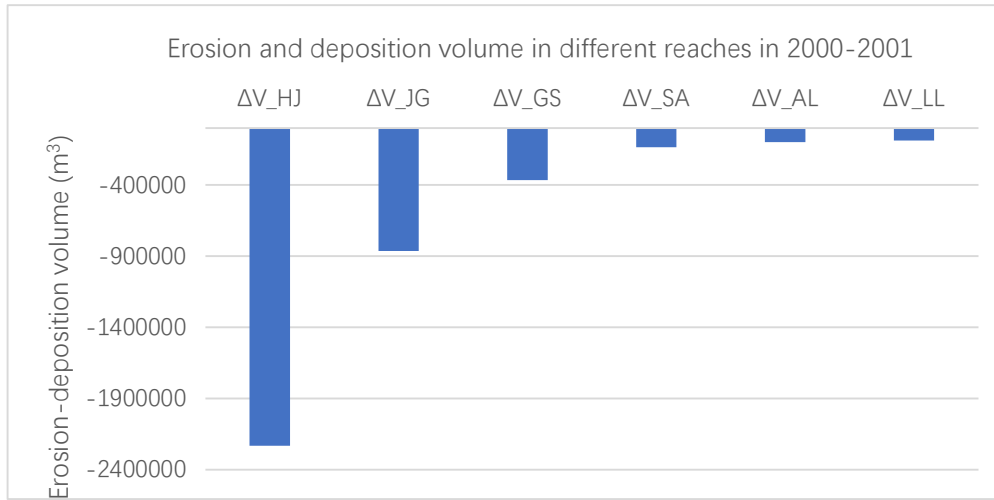

Figure 4 Total Scour-and-Silt Volume by River Reach, 2000-2001

**4 Discussion and Conclusion**

The calculations show that the monthly number of sediment particles transported in the studied reaches during 2000-2001 all exceeded $5 \times 10^{19}$. Figure 2 reveals a seasonal variation in the number of particles participating in transport in the Lower Yellow River below Xiaolangdi: higher in winter and lower in summer. In Figure 3, values below 0 represent scour, and values above 0 represent siltation. During these 24 months, scour occurred from January to July, with a total scour volume on the order of tens of millions of cubic meters. Siltation occurred from July to the following January, with a total siltation volume also on the order of tens of millions of cubic meters. Scour and siltation alternated, showing a rough "balance," with the total scour volume exceeding the total siltation volume. Comparing the reaches, the HJ reach had the largest scour-and-silt volumes, followed by the JG reach. The scour-and-silt volumes gradually decreased from upstream to downstream, with a noticeable decrease from Huayuankou to Sunkou, and a slow decrease downstream from Sunkou. Figure 4 shows that overall, the studied reaches experienced net scour in 2000-2001, with the total scour volume decreasing from upstream to downstream.

According to the Yellow River Sediment Bulletin 2001 (Hydrology Bureau, Yellow River Conservancy Commission, 2002), the sediment discharge (in tons) for Huayuankou-Lijin was generally higher in spring and autumn and lower in winter and summer. The variation pattern of the number of sediment particles in this calculation partially coincides with this in summer. The number of particles participating in transport was higher in winter, but the sediment discharge mass was lower. Based on available data, measurement errors cannot be estimated. Simultaneously, the sediment gradation and dry bulk density used in this calculation are multi-year averages; there are no measured data for 2000 and 2001 specifically, nor seasonal gradation data. Consequently, the derived average particle mass affects the calculation of particle numbers, as the mass of sediment particles actually participating in transport is influenced by seasonal variations in upstream water and sediment inflow and reservoir regulation. The calculated average particle mass used here cannot reflect this seasonal variation, but the overall pattern is consistent with observations in the Lower Yellow River after the operation of the Xiaolangdi Reservoir (Ma,et al, 2024; Zhang et al, 2025). If observational errors are not large enough to affect the overall pattern, it is plausible that during winter with lower flow rates, the transported particles are likely smaller in size and lighter in mass, leading to a higher number of sediment particles participating in transport but a lower total sediment discharge mass.

Existing long-term observations of the Lower Yellow River reaches found persistent siltation before 2000, with the siltation volume gradually decreasing from upstream to downstream. Starting from the operation of the Xiaolangdi Reservoir in 2000, the trend shifted to overall scour, with the scour volume decreasing from upstream to downstream, and the flood season changing from siltation to scour (Li et al, 2024; Ma et al, 2024; Zhang et al, 2025). The results of this calculation align well with these observations. However, it should be noted that there is room for improvement in the process of deriving the sediment bulk volume from the particle volume using the ratio $\phi_{RCP}$.

The method adopted in this study is based on extensive simulation experiments and accounts for the influence of particle size distribution polydispersity and skewness (Desmond & Weeks, 2014). The original experiments did not involve sediment particles with as many size fractions as those in the research region (see Table 1), which may have certain implications for the results. Moreover, other particle packing experiments, based on different particle generation mechanisms, propose alternative approaches for calculating $\phi_{RCP}$. Currently, the bulk volume is derived from simulation experiments, which typically aim to determine the maximum possible volume fraction of particles of varying sizes within a confined space of given volume, e.g., a cylinder or cube. This differs from the context of a river channel, which is an unconfined space. Therefore, the original definition of $\phi_{RCP}$ has been extended here to represent the volume fraction per unit bulk volume. In natural riverbeds, sediment particles can move in various directions, meaning that the unit volume can effectively "overflow." This may also influence the results. Although the extended $\phi_{RCP}$ has yielded reasonably realistic estimates of sediment bulk volume in the study river reach, there remains potential for refinement. Future work could develop more suitable expressions based on simulations that better reflect natural conditions. However, such efforts lie beyond the scope of this study. Here, we only discuss the potential implications of the method used to obtain $\phi_{RCP}$ on the estimation of sediment bulk volume.

In summary, the macroscopic physical model of sediment movement established in this study, by treating the river channel as an isothermal-isobaric ensemble containing a large number of non-identical moving particles, provides calculations for the number of sediment particles participating in transport and the scour-and-silt volumes along the Huayuankou-Lijin reach of the Lower Yellow River during 2000-2001 that are relatively consistent with observational facts. Scour-and-silt changes in wide, shallow rivers can be simulated using a statistical physics model that requires no empirical parameters. This method will be applied in subsequent work to simulate river scour and siltation in more diverse environments. Since the required independent variables are few, all are observable, and feasible alternative calculation methods exist for these quantities in the absence of observations, widespread application is facilitated. Furthermore, as the ideal model of this method is the movement of multi-scaled, large numbers of particles in a general isothermal-isobaric ensemble, it may not only be applicable to river channels but also potentially to other multi-particle systems satisfying the conditions, which awaits further research and verification.

**Funding**
This work is supported by the Open Research Fund Program of the State Key Laboratory of Hydroscience Science and Engineering (Application of a theorized delayed response dynamic model in the construction of a digital 3-D river network, sklhse-2024-B-01).

**Availability of data and materials**
All the available data were mentioned within the manuscript.